\documentclass[amsmath,amssymb,showpacs,draft,preprint,floatfix]{revtex4}
\usepackage{graphicx}
\begin{document}

\title[P. C. Garc\'{i}a Quijas, et. al.]{Factorizing the time evolution operator }

\author{P. C. Garc\'{i}a Quijas}
\author{L. M. Ar\'{e}valo Aguilar\footnote[1]{larevalo@cio.mx} }
\address{Centro de Investigaciones en \'Optica, Loma del Bosque No. 115,
Fracc. Lomas del Campestre, Le\'{o}n, Gto.  M\'{e}xico.}

\begin{abstract}
There is a widespread belief in the quantum physical community,
and in textbooks used to teach Quantum Mechanics,  that it is a
difficult task to apply the time evolution operator
$e^{it\hat{H}/\hbar}$ on an initial wave function. That is to say,
because the hamiltonian operator generally is the sum of two
operators, then it is a difficult task to apply the time evolution
operator on an initial wave function $\psi(x,0)$, for it implies
to apply terms like $(\hat{a}+\hat{b})^n$. A possible solution of
this problem is to factorize the time evolution operator and then
apply successively the individual exponential operator on the
initial wave function. However, the exponential operator does not
directly factorize, i. e. $e^{\hat{a}+\hat{b}}\neq
e^{\hat{a}}e^{\hat{b}}$. In this work we present a useful
procedure for factorizing the time evolution operator when the
argument of the exponential is a sum of two operators, which obey
specific commutation relations. Then, we apply the exponential
operator as an evolution operator for the case of elementary
unidimensional potentials, like the particle subject to a constant
force and the harmonic oscillator. Also, we argue about an
apparent paradox concerning the time evolution operator and
non-spreading wave packets addressed previously in the literature.
\end{abstract}


\maketitle

\section{Introduction}

Quantum Mechanics is a successful theory. Although highly
counterintuitive, using it we are able to explain the microscopic
world. Also, Quantum Mechanics have discovered many natural
process which have culminated in practical technological
applications, like the transistor and the Quantum Cryptography.
However, Quantum Mechanics is a difficult field of study. For it
takes many years to develop the necessary skills to understand
their relevant concepts.

One of the hardest skills to develop is to understand the
technique to solve the fundamental equation of Quantum Mechanics,
i. e. the Schr\"{o}dinger equation. In fact, there are few cases
where this equation was analytically solved. One of the principal
factors that impede the straightforward solution is that this
equation involves not usual mathematical concepts, like operators:

\begin{equation}
i\hbar \frac{\partial}{\partial
t}\Psi(t)=\hat{H}\Psi(t),\label{schrodinger}
\end{equation}
where the Hamiltonian $\hat{H}$ is an operator that has to be
self-adjoint, and it is generally the sum of two or more
operators, let us say:
\begin{equation}
\hat{H}=\hat{a}+\hat{b}.\label{suma1}
\end{equation}

In the literature, the most used technique to solve Equation
(\ref{schrodinger}) is to find the eigenvalues and eigenfunctions
of the time independent Schr\"{o}dinger equation:
\begin{equation}
\hat{H}\psi_n(x)=E_n\psi_n(x)\label{eigenvalueEq},
\end{equation}
where $E_n$ and $\psi_n(x)$ are, respectively, the eigenvalues and
eigenfunctions of the hamiltonian $\hat{H}$
\cite{book1,book2,book3,book4}. Then, the time dependent wave
function is constructed taking the superposition of the
eigenfunctions of the hamiltonian:
\begin{equation}
\Psi(x,t)=\sum_{n=0}^{\infty}e^{-iE_nt/\hbar}c_n\psi_n(x)\label{eigenvalueEq},
\end{equation}
where $c_n=\int \Psi(x,0) \psi_n dx$ is the scalar product between
the initial state of the system and the eigenfuctions of the
hamiltonian. In this paper, we call this method \textit{the
eigenstates method}.

A second way for solving Eq. (\ref{schrodinger}), if we consider a
time independent Hamiltonian, is to integrate Eq.
(\ref{schrodinger}) with respect to time, to obtain \cite{book1}:
\begin{equation}
\Psi(x,t)=e^{-\frac{it}{\hbar}\hat{H}}\Psi(x,0)=e^{\hat{A}+\hat{B}}\Psi(x,0),\label{sol1}
\end{equation}
where $\Psi(x,0)$ is the initial wave vector,
$\hat{A}=-(it/\hbar)\hat{a}$ and $\hat{B}=-(it/\hbar)\hat{b}$. In
this paper, we call this method \textit{the evolution operator
method}.

Essentially, both ways for solving the Schr\"{o}dinger equation
are the same. This can be proved by expanding $\Psi(x,0)$ in terms
of the eigenfunctions of the Hamiltonian, i.e. $\Psi(x,0)=\sum
c_n\psi_n(x)$, and inserting it on the right hand side of Eq.
(\ref{sol1}) to produce Equation (\ref{eigenvalueEq}). An
alternative way to solve the Schr\"{o}dinger equation is the
technique developed by Feynman, called the Feynman propagator
method \cite{holstein2,gori,barone}.

The trouble with Equation (\ref{sol1}) is that, in general,
$\hat{A}$ and $\hat{B}$ do not commute. This makes difficult to
apply the time evolution operator to the initial state vector
given in Eq. (\ref{sol1}). In fact, the problem is how to make the
expansion of a function of noncommuting operators like that in Eq.
(\ref{sol1}), i. e.
$e^{\hat{A}+\hat{B}}=\sum_{n=0}^{\infty}(1/n!)(\hat{A}+\hat{B})^n$,
in such a way that all the $\hat{B}$ precede the $\hat{A}$, or
viceversa. This problem has been already studied by many authors,
and some theorems have been proved to handle this expansion. For
example, Kumar proved the following expansion for a function of
noncommuting operators \cite{kumar}:
\begin{equation}
f(\hat{A}+\hat{B})=\sum_{n=0}^{\infty}\frac{1}{n!}C^n(\hat{A},\hat{B})f^{(n)}(\hat{A}),\label{equ-kumar}
\end{equation}
where $C^n(\hat{A},\hat{B})$ is a coefficient operator given in
terms of $\hat{B}$ and the commutator $[\hat{A},\hat{B}]$
\cite{kumar}.

Also, Cohen has proved the following expansion theorem for the
operators $\hat{x}$ and $\hat{p}$ \cite{cohen}: Given a function
$F(\hat{x},\hat{p})$ then
\begin{equation}
F^n(\hat{x},\hat{p})=\sum_{n=0}^\infty\alpha_k^nu_k(\hat{x})\int_{-\infty}^\infty
u_k^*(\hat{x}+\theta)e^{(i/\hbar)\theta
p}d\theta,\label{equ-cohen}
\end{equation}
where $\alpha_k$ and $u_k(\hat{x})$ are the eigenvalue and
eigenfunction of the eigenvalue problem
$F(\hat{x},\hat{p})u_k(\hat{x})=\alpha_ku_k(\hat{x})$. In
particular, the expansion for the function
$(\lambda\hat{x}+\hat{p})^n$ has given as \cite{cohen}:
\begin{equation}
(\lambda\hat{x}+\hat{p})^n=\sum_{k=0}^{[\frac{1}{2}n]}\sum_{s=0}^{n-2k}\frac{(-1)^kn!}{k!(n-2k)!}
\left(\begin{array}{cc}n-2k\\s\end{array} \right)
(i\hbar/2)^k\lambda^{n-k-s}
\hat{x}^{n-2k-s}\hat{p}^s.\label{exp-cohen}
\end{equation}
In general, these expansion theorems have produced a high
cumbersome expressions that are very difficult to apply.

One of the possible paths to avoid the expansion of functions of
two noncommuting operators, in the case of the exponential
operators, is to factorize the argument of the exponential. This
approach facilitates the application of the exponential operator
because now, when the exponential operator is factorized, we have
only to expand the exponential of a single operator, i. e.
$e^{\hat{A}}e^{\hat{B}}$, which is more simple. However, the
factorization of exponential operators is not an easy task. To our
best knowledge, the evolution operator method has been applied to
unidimensional problems in only four other related articles
\cite{blinder,robinett,bala,cheng}.

The main goal of this paper is twofold, first we shall show a
procedure to factorize the exponential operator and, secondly, we
shall show how to apply the factorized exponential operator on an
initial wave function. The method of factorization that we shall
present in this paper has been used in Quantum Optics. Also, this
method has been proposed as a possible tool to improve some
misconceptions in the teaching of Quantum Mechanics \cite{paulo}.
Therefore, an important objective of this paper is to review this
method in order that it becomes available for the people outside
these fields.

Although the three methods for solving the Schr\"{o}dinger
equation mentioned above have to give the same result, the
evolution operator method is, in some way, quite different of the
eigenstates method and the Feynman propagator. For example, in the
eigenstates method we need to look for the eigenfunctions of the
hamiltonian where the particle is placed; on the contrary, the
evolution operator method does not give any information about the
eigenfunctions of the hamiltonian. Also, the Feynman propagator
method need to look for all the possible paths the particle can
take from an initial wave function to a final one, and the
evolution operator method does not inquire for these possible
paths. In some sense, the evolution operator method is more direct
than the other two.

In summary, this paper address the problem of how to factorize the
exponential of a sum of operators, in order to be able to apply it
as an evolution operator, when the operators obey certain
commutation rules. To make the factorization we use the tool of
the differential equation method \cite{wilcox,louisell}, which
requires that both sides of an equation satisfy the same
first-order differential equation and the same initial condition.
For a review of these tools see the work of Wilcox \cite{wilcox}
and Lutzky \cite{lutzky}. This method has been used successfully
in the field of Quantum Optics \cite{yo,yo2,lu1}. We shall show
that this method is useful and easy to apply in the unidimensional
problems of Quantum Mechanics.

This paper is organized as follows: In Section II we will present
the method and show how to apply it for factorizing an exponential
operator. In Section III we give a specific example when the
operators obey certain commutation rules. In Subsection A of this
section, we apply the found factorization to the case when the
particle is subjected to a constant force. In Section IV we
present the factorization of the exponential operator when its
argument obeys a more complex commutation rules; in subsection A
of this section the factorization found is applied to the harmonic
oscillator. In subsection B of section IV, we derive another way
to factorize the harmonic oscillator and we show that both
factorizations give the same evolution function (In Appendix A we
derive yet another way to factorize the harmonic oscillator). In
Section V we address a supposed limitation of the evolution
operator method, we demonstrate that the limitation is because the
initial wave function used to show the apparent paradox is outside
of the domain of the hamiltonian operator.

\section{The method}

As the global purpose of this paper is pedagogical, in this
section we show how the method works. Our intention is that this
method can be used for researchers of any field to find the
evolution state from an initial wave function. In order to be
explicit we separate the method in three steps and apply it to
obtain the well know Baker-Cambell-Hausdorff formula:
\begin{eqnarray}
e^{(\hat{A}+\hat{B})}=e^{\hat{A}}e^{\hat{B}}e^{f([\hat{A},\hat{B}])}.\qquad \quad\\
\quad When
\quad[\hat{A},[\hat{A},\hat{B}]]=[\hat{B},[\hat{A},\hat{B}]]=0
\quad i. e. \quad [\hat{A},\hat{B}]=\delta. \nonumber
\end{eqnarray}
where $\delta$ is a constant. Notice that after we have solved
this easy problem we will progressively increase the difficulty of
the commutation relation.

 To make the factorization of the exponential of the sum of two operators we proceed as follows:

\begin{enumerate}
\item Firstly, we define an auxiliary function in terms of the
exponential of the sum of two operators, its commutator and an
auxiliary parameter $\xi$:
\begin{equation}
F(\xi)=e^{\xi(\hat{A}+\hat{B})}, \label{F-definition1}
\end{equation}

\begin{equation}
F(\xi)=e^{f_1(\xi)[\hat{A},\hat{B}]}e^{f_2({\xi})\hat{A}}
e^{f_3({\xi})\hat{B}}. \label{F-definition2}
\end{equation}

Note that in Equations (\ref{F-definition1}) and
(\ref{F-definition2}) we are defining separately $F(\xi)$ as a
function and its factorization. That is to say, if
$e^{f_1(\xi)[\hat{A},\hat{B}]}e^{f_2({\xi})\hat{A}}
e^{f_3({\xi})\hat{B}}$ is the factorization of
$e^{\xi(\hat{A}+\hat{B})}$, then
$F(\xi)=e^{\xi(\hat{A}+\hat{B})}=e^{f_1(\xi)[\hat{A},\hat{B}]}e^{f_2({\xi})\hat{A}}
e^{f_3({\xi})\hat{B}}$.

\item Secondly, we differentiate Equations (\ref{F-definition1})
and (\ref{F-definition2}) with respect to the parameter $\xi$ to
obtain:
\begin{eqnarray}
\frac{dF(\xi)}{d\xi}=(\hat{A}+\hat{B})e^{\xi(\hat{A}+\hat{B})}=(\hat{A}+\hat{B})F(\xi),
\label{derivada}
\end{eqnarray}

\begin{eqnarray}
\frac{dF(\xi)}{d\xi}=
\frac{df_1(\xi)}{d\xi}[\hat{A},\hat{B}]e^{f_1(\xi)[\hat{A},\hat{B}]}e^{f_2({\xi})\hat{A}}
e^{f_3({\xi})\hat{B}}+ \nonumber\\
e^{f_1(\xi)[\hat{A},\hat{B}]}\frac{df_2({\xi})}{d\xi}\hat{A}e^{f_2({\xi})\hat{A}}
e^{f_3({\xi})\hat{B}}+e^{f_1(\xi)[\hat{A},\hat{B}]}e^{f_2({\xi})\hat{A}}
\frac{df_3({\xi})}{d\xi}\hat{B}e^{f_3({\xi})\hat{B}}.
\label{derivada2}
\end{eqnarray}
After that, we need to put in order the operators of Equation
(\ref{derivada2}). In order to make the new arrange we use the
fact that the operators are self-adjoints, i. e.
$e^{\xi\hat{A}}\hat{B}=\left(e^{\xi\hat{A}}\hat{B}e^{-\xi\hat{A}}\right)e^{\xi\hat{A}}$,
and the well know relation:
$e^{\gamma\hat{A}}\hat{B}e^{-\gamma\hat{A}}=\hat{B}+\gamma
[\hat{A},\hat{B}]+\frac{\gamma^2}{2!}[\hat{A},[\hat{A},\hat{B}]]+\dots$,
see reference \cite{louisell}. That is, we have to pass the
exponentials to the right in the right hand side of Equation
(\ref{derivada2}).  In this case:

\begin{equation}
e^{f_1(\xi)[\hat{A},\hat{B}]}\hat{A}e^{-f_1(\xi)[\hat{A},\hat{B}]}=
\hat{A}
\end{equation}
 and
\begin{equation}
e^{f_2(\xi)\hat{A}}\hat{B}e^{-f_2(\xi)\hat{A}}=\hat{B}+\delta
f_2(\xi),
\end{equation}
where we have used the fact that $[\hat{A},\hat{B}]=\delta$. If we
substitute these relations in Eq. (\ref{derivada2}) we obtain:
\begin{eqnarray}
\frac{dF(\xi)}{d\xi}=
\frac{df_1(\xi)}{d\xi}[\hat{A},\hat{B}]e^{f_1(\xi)[\hat{A},\hat{B}]}e^{f_2({\xi})\hat{A}}e^{f_3({\xi})\hat{B}}+ \nonumber\\
\frac{df_2({\xi})}{d\xi}\hat{A}e^{f_1(\xi)[\hat{A},\hat{B}]}e^{f_2({\xi})\hat{A}}
e^{f_3({\xi})\hat{B}}+ \nonumber
\\\frac{df_3({\xi})}{d\xi}e^{f_1(\xi)[\hat{A},\hat{B}]}
[\hat{B}+\delta
f_2(\xi)]e^{f_2({\xi})\hat{A}}e^{f_3({\xi})\hat{B}}.
\label{arreglo1}
\end{eqnarray}

Now, using the relation
$e^{f_1(\xi)[\hat{A},\hat{B}]}\hat{B}e^{-f_1(\xi)[\hat{A},\hat{B}]}=\hat{B}$
in Eq. (\ref{arreglo1}) we obtain:
\begin{eqnarray}
\frac{dF(\xi)}{d\xi}=
\frac{df_1(\xi)}{d\xi}[\hat{A},\hat{B}]e^{f_1(\xi)[\hat{A},\hat{B}]}e^{f_2({\xi})\hat{A}}e^{f_3({\xi})\hat{B}}+ \nonumber\\
\frac{df_2({\xi})}{d\xi}\hat{A}e^{f_1(\xi)[\hat{A},\hat{B}]}e^{f_2({\xi})\hat{A}}
e^{f_3({\xi})\hat{B}}+ \nonumber
\\\frac{df_3({\xi})}{d\xi}
[\hat{B}+\delta
f_2(\xi)]e^{f_1(\xi)[\hat{A},\hat{B}]}e^{f_2({\xi})\hat{A}}e^{f_3({\xi})\hat{B}}.
\label{arreglo2}
\end{eqnarray}

That is, we successfully passed all the exponential to the right
and we can write Eq. (\ref{arreglo2}) as:
\begin{eqnarray}
\frac{dF(\xi)}{d\xi}=
\left\{\frac{df_1(\xi)}{d\xi}[\hat{A},\hat{B}]+
\frac{df_2({\xi})}{d\xi}\hat{A}+ \frac{df_3({\xi})}{d\xi}
[\hat{B}+\delta f_2(\xi)] \right\} \nonumber
\\ \times e^{f_1(\xi)[\hat{A},\hat{B}]}e^{f_2({\xi})\hat{A}}e^{f_3({\xi})\hat{B}}.
\label{arreglo3}
\end{eqnarray}
That is, by Equation (\ref{F-definition2}) we arrive to the
following result:
\begin{eqnarray}
\frac{dF(\xi)}{d\xi}=
\left\{\frac{df_1(\xi)}{d\xi}[\hat{A},\hat{B}]+
\frac{df_2({\xi})}{d\xi}\hat{A}+ \frac{df_3({\xi})}{d\xi}
[\hat{B}+\delta f_2(\xi)] \right\} F(\xi). \label{arreglo4}
\end{eqnarray}

\item Finally, as a third step, we  must to compare the
coefficients of Eq. (\ref{derivada}) and Eq. (\ref{arreglo4}),
from which a set of differential equations is obtained:
\begin{eqnarray}
\frac{df_2(\xi)}{d\xi}&=&1, \nonumber \\
\frac{df_3(\xi)}{d\xi}&=&1, \nonumber \\
\delta\frac{df_1(\xi)}{d\xi}+\delta
\frac{df_3(\xi)}{d\xi}f_2(\xi)&=&0,
\end{eqnarray}
subjected to the initial condition $f_1(0)=f_2(0)=f_3(0)=0$. In
this case the solutions are:
\begin{eqnarray}
f_2(\xi)&=&\xi, \nonumber \\
f_3(\xi)&=&\xi, \nonumber \\
f_1(\xi)&=&-\frac{1}{2}\xi^2. \label{sol-metodo}
\end{eqnarray}
After substituting Eq. (\ref{sol-metodo}) in Eq.
(\ref{F-definition2}) we arrive to the following Equation:
\begin{equation}
e^{\xi(\hat{A}+\hat{B})}=e^{-\xi^2[\hat{A},\hat{B}]/2}e^{\xi\hat{A}}e^{\xi\hat{B}}.
\end{equation}
Setting $\xi=1$ we obtain the usual Baker-Campbell-Hausdorff
formula.
\end{enumerate}

This method facilitates the application of the exponential
operator, because now we have to handle only individual operators
function. The proposed factorization of Equation
(\ref{F-definition2}) is one of the possibilities, also we can
define $F(\xi)$ as:
\begin{equation}
F(\xi)=
e^{g_1({\xi})\hat{B}}e^{g_2({\xi})\hat{A}}e^{g_3(\xi)\hat{B}},\label{alternative}
\end{equation}
or make another arrange of the exponentials, as for example
$e^{h_1({\xi})\hat{A}}e^{h_2({\xi})\hat{B}}e^{h_3(\xi)\hat{A}}$.
Notice that Equation (\ref{alternative}), in contrast to Equation
(\ref{F-definition2}), does not use the commutator in the
exponential functions. This arrangement could be used to treat
specific problems, as the harmonic oscillator.

In fact, when the method is dominated this arrangement is a set of
crafted directions, which gives a factorization of the evolution
operator. In the majority of cases, a different arrange will
produce a different set of differential equations and, obviously,
a different set of solutions. We give an explicit example of this
fact in the case of the harmonic oscillator, see Equations
(\ref{x-p-x-fac}), (\ref{funct-oscillator2}) and
(\ref{funct-oscilator}). It is very important not to confuse the
Baker-Campbell-Hausdorff formula with the method addressed here.
Each one represents a different way to factorize exponential
operators.

On the other hand, this method can be used to improve some
misconceptions of Quantum Mechanics \cite{paulo,singh2}.
Immediately we present the different cases that appear when the
operators obey different commutation rules.

\section{Case 1: $[\hat{A},\hat{B}]=\hat{C},\quad[\hat{A},\hat{C}]=0$ and $[\hat{C},\hat{B}]=-k$}

This section is organize as follows: Firstly, we make the
factorization of the exponential operator when the operators obey
the commutation relations given by Equation (\ref{commutation1}).
Secondly, in Subsection A we use the factorized exponential to
solve the problem of a particle subjected to a constant force.

 Therefore, we begin the factorization of
exponential operators by analyzing the case when
\begin{equation}
[\hat{A},\hat{B}]=\hat{C},\quad[\hat{A},\hat{C}]=0,\quad and \quad
[\hat{C},\hat{B}]=-k,\label{commutation1}
\end{equation}
where $\hat{A}$, $\hat{B}$ and $\hat{C}$ are operators and $k$ is
a c-number (in general, we use the simbol $\hat{}$ to denote
operators).

In the present case, we propose the factorization function as:
\begin{equation}
F(\xi)=e^{\xi(\hat{A}+\hat{B})}=e^{f(\xi)\hat{A}}e^{g(\xi)\hat{B}}e^{h(\xi)\hat{C}}e^{r(\xi)},\label{f1}
\end{equation}
by differentiating Eq. (\ref{f1}) with respect to $\xi$ we obtain
for the left hand side
\begin{equation}
\frac{dF(\xi)}{d\xi}=(\hat{A}+\hat{B})F(\xi),\label{dif1a}
\end{equation}
and, for the right hand side
\begin{equation}
\frac{dF(\xi)}{d\xi}=\left[
\frac{df}{d\xi}\hat{A}+\frac{dg}{d\xi}\hat{B}+\frac{dg}{d\xi}f(\xi)\hat{C}+\frac{dh}{d\xi}(\hat{C}+kg)+\frac{dr}{d\xi}
\right]F(\xi),\label{dif1b}
\end{equation}
where we have applied the fact that
\begin{equation}
e^{\xi\hat{A}}\hat{B}e^{-\xi\hat{A}}=\hat{B}+\xi[\hat{A},\hat{B}]+\frac{\xi^2}{2!}
[\hat{A},[\hat{A},\hat{B}]]+\ldots,\label{BCH}
\end{equation}
and we have used the commutation relation of Eq.
(\ref{commutation1}).

By equating the coefficients of Eq. (\ref{dif1a}) and Eq.
(\ref{dif1b}) we obtain the following system of differential
equations:
\begin{eqnarray}
\frac{df(\xi)}{d\xi}&=&1,\nonumber\\
\frac{dg(\xi)}{d\xi}&=&1,\nonumber \\
\frac{dg(\xi)}{d\xi}f(\xi)+\frac{dh(\xi)}{d\xi}&=&0,\nonumber\\
kg(\xi)\frac{dh(\xi)}{d\xi}+\frac{dr(\xi)}{d\xi}&=&0\label{set1}
\end{eqnarray}
subjected to the initial condition $F(0)=0$, which implies:
\begin{equation}
f(0)=g(0)=h(0)=r(0)=0.\label{condition1}
\end{equation}

By solving Eq. (\ref{set1}) with the initial condition stated in
Eq. (\ref{condition1}), we finally obtain
\begin{equation}
e^{\xi(\hat{A}+\hat{B})}=e^{(\xi^3/3)k}e^{\xi\hat{A}}e^{\xi\hat{B}}e^{-(\xi^2/2)\hat{C}}.\label{fact1}
\end{equation}
Setting $\xi=1$ we obtain the factorization we were looking for.

\subsection{Application: A particle subject to a constant force}
One application of the evolution operator method is when we study
the time dependence of a quantum state. There have been some
results in this approach when the operator is the energy of a free
particle \cite{blinder}, or the energy of a particle subjected to
a constant force, that is $V(x)=-Fx$ \cite{robinett}. In this
subsection, we use the factorization found above to solve the
problem of a particle subject to a constant force, whit the help
of the Blinder's method \cite{blinder}. The Blinder's method show
how to apply the evolution operator like and infinite sum for a
free particle:
\begin{equation}
e^{\frac{-it\hat{p}^2}{2m\hbar}}=\sum_{n=0}^\infty\left(\frac{i\hbar
t}{2m}\right)^n\frac{1}{n!} \left( \frac{\partial^2}{\partial
x^2}\right)^n. \label{evol-sum}
\end{equation}
For a free particle the wave function at time $t$ is obtained by
operating with the evolution operator on the initial wave
function. Taking as an initial wave function:
\begin{equation}
\Psi(x,0)=\frac{1}{(\sigma\sqrt{\pi})^{1/2}}e^{-\frac{x^2}{4\sigma^2}},\label{initial-wave}
\end{equation}
where $\sigma$ is the width of the wave packet. The Blinder's
method consist in the application of the identity \cite{blinder}:
$\frac{\partial^2}{\partial x^2}\left\{
z^{-1/2}\exp\{-x^2/4z\}\right\}=\frac{\partial}{\partial z}\left\{
z^{-1/2}\exp\{-x^2/4z\}\right\}$. This identity allows us to apply
the evolution operator on initial wave functions like gaussian
wave packets, for details see reference \cite{blinder}.

For a particle subject to a constant force, i.e. $V(x)=-Fx$, the
wave function at time $t$ is given by:
\begin{equation}
\Psi(x,t)=\exp\left[-\frac{it}{\hbar}
\left(\frac{\hat{p}^2}{2m}-Fx \right)
\right]\Psi(x,0),\label{ini-ecuatio}
\end{equation}
defining $\hat{A}=\hat{p}^2(2m)^{-1}$, $\hat{B}=-F\hat{x}$, and
using the commutation relations between $\hat{p}$ and $\hat{x}$ we
can deduce the following commutation rules:
\begin{equation}
[\hat{A},\hat{B}]=\frac{i\hbar F}{m}\hat{p},\qquad
[\hat{A},\hat{C}]=0,\qquad and \qquad
[\hat{C},\hat{B}]=-\frac{\hbar^2F^2}{m},\label{commutation3}
\end{equation}
where $\hat{C}=i\hbar F\hat{p}/m$. If we identify
$k=\hbar^2F^2/m$, then the commutation relations of Eq.
(\ref{commutation3}) are similar to that of Eq.
(\ref{commutation1}). Therefore, if we use Eq. (\ref{fact1}), we
can write Eq. (\ref{ini-ecuatio}) as:
\begin{equation}
\Psi(x,t)=\exp\left[\frac{it^3F^2}{3\hbar
m}\right]\exp\left[-\frac{it}{2m\hbar}\hat{p}^2\right]\exp\left[\frac{itF}{\hbar}\hat{x}\right]
\exp\left[\frac{it^2F}{2m\hbar}\hat{p}\right]\Psi(x,0).\label{evol-fx1}
\end{equation}

Using the theorem
$\exp[\xi\hat{A}]F(\hat{B})\exp[-\xi\hat{A}]=F(\exp[\xi\hat{A}]\hat{B}\exp[-\xi\hat{A}])$
\cite{louisell}, we can rearrange Eq. (\ref{evol-fx1}) as follows:
\begin{equation}
\Psi(x,t)=\exp\left[-\frac{iF^2t^3}{6m\hbar
}\right]\exp\left[\frac{itF}{\hbar}\hat{x}\right]\exp\left[-\frac{it}{2m\hbar}\hat{p}^2\right]
\exp\left[-\frac{it^2F}{2m\hbar}\hat{p}\right]\Psi(x,0).\label{evol-fx2}
\end{equation}
Taking as an initial state that of Eq. (\ref{initial-wave}) we
finally obtain:
\begin{eqnarray}
\Psi(x,t)=\left(\sigma\sqrt{2\pi}\right)^{-1/2}\left(1+(i\hbar
t/2m\sigma^2)\right)^{-1}\exp\left[\frac{iFt}{\hbar}\left(x-(Ft^2/6m)\right)\right]
\nonumber
\\ \times\exp\left[-\frac{(x-\left(Ft^2/2m)\right)^2}{4\left(\sigma^2+(i\hbar
t/2m)\right)}\right].\label{final1}
\end{eqnarray}
Equations (\ref{evol-fx2}) and (\ref{final1}) are exactly the same
equations obtained by Robinett \cite{robinett}.

\section{Case 2: $[\hat{A},\hat{B}]=\hat{C},\quad \lbrack \hat{A},\hat{C}%
]=2\protect\gamma \hat{A}$ and $[\hat{B},\hat{C}]=-2\protect\gamma
\hat{B}$}

In this section, we carry out the factorization of the exponential
operator when the commutation rules are given by Equation
(\ref{commu-oscillator}). Then, we will show in Subsection A that
these commutation relations are the same of the harmonic
oscillator. On the other hand, in Subsection B we show an
alternative way of factorization for this problem, and show that
the evolution given by the evolution operator are the same in both
cases.

\begin{equation}
\lbrack \hat{A},\hat{B}]=\hat{C},\quad \lbrack \hat{A},\hat{C}]=2\gamma\hat{%
A} \quad [\hat{B},\hat{C}]=-2\gamma \hat{B}.
\label{commu-oscillator}
\end{equation}
In this case, using an arrangement similar to Equation
(\ref{alternative}), we define the function as:
\begin{equation}
F(\xi )=e^{\xi (\hat{A}+\hat{B})}=e^{f(\xi )\hat{B}}e^{g(\xi )\hat{A}%
}e^{h(\xi )\hat{B}}. \label{p-x-p}
\end{equation}
By differentiating Equation (\ref{p-x-p}) with respect to $\xi $,
we obtain:
\begin{equation}
 \frac{dF(\xi )}{d\xi
}=(\hat{A}+\hat{B})F(\xi ), \label{der1-p-x-p}\end{equation}
 and
\begin{eqnarray}
\frac{dF(\xi )}{d\xi }=\Bigg\{ \left[g(\xi) \frac{dh(\xi)}{d\xi
}-f(\xi)\frac{dg(\xi)}{d\xi}-\gamma
g^2(\xi)f(\xi)\frac{dh(\xi)}{d\xi} \right] \hat{C} \nonumber\\
\left[ \frac{df(\xi)}{d\xi}+\gamma f^2(\xi)
\frac{dg(\xi)}{d\xi}+\frac{dh(\xi)}{d\xi}-2\gamma f(\xi) g(\xi)
\frac{dh(\xi)}{d\xi}+\gamma^2 f^2(\xi) g^2(\xi)
\frac{dh(\xi)}{d\xi}\right]\hat{B}
 \nonumber \\
+\left[ \frac{dg(\xi)}{d\xi }+\gamma g^{2}(\xi)\frac{dh(\xi)}{d\xi
}\right] \hat{A}\Bigg\} F(\xi ),\label{der-p-x-p}
\end{eqnarray}
where we have applied the relation $
e^{\xi \hat{A}}\hat{B}e^{-\xi \hat{A}}=\hat{B}+\xi \lbrack \hat{A},\hat{B}]+%
\frac{\xi ^{2}}{2!}[\hat{A},[\hat{A},\hat{B}]]+\ldots . $

By equating Equations (\ref{der1-p-x-p}) and (\ref{der-p-x-p}), we
obtain the following system of differential equations:

\begin{eqnarray}
\frac{dg(\xi)}{d\xi }+\gamma g^{2}(\xi)\frac{dh(\xi)}{d\xi
} =1, & \nonumber \\
\frac{df(\xi)}{d\xi}+\gamma f^2(\xi)
\frac{dg(\xi)}{d\xi}+\frac{dh(\xi)}{d\xi}-2\gamma f(\xi) g(\xi)
\frac{dh(\xi)}{d\xi}+\gamma^2 f^2(\xi) g^2(\xi)
\frac{dh(\xi)}{d\xi}=1, & \nonumber \\
g(\xi) \frac{dh(\xi)}{d\xi }-f(\xi)\frac{dg(\xi)}{d\xi}-\gamma
g^2(\xi)f(\xi)\frac{dh(\xi)}{d\xi}=0, & \label{difer-p-x-p}
\end{eqnarray}
subjected to the initial condition $F(0)=1$, which means:
$g(0)=f(0)=h(0)=0.
$
By solving Equation (\ref{difer-p-x-p}), with the initial
conditions, we obtain the following solutions:

\begin{eqnarray}
f\left( \xi \right) &=&h\left( \xi \right)
=\frac{1}{\sqrt{\gamma}}\tan \left( \xi \sqrt{\gamma }/2\right),  \\
g\left( \xi \right)&=&\frac{1}{\sqrt{\gamma }}\sin \left( \xi
\sqrt{\gamma } \right).
\end{eqnarray}
Setting $\xi=1$ we obtain the factorization we were looking for,
that is Eq. (\ref{p-x-p}).

As it was stated at the end of Section $2$, the factorization
given in Equation (\ref{p-x-p}) is only one of many possibilities.
Since the operators do not commute, various orderings on the right
hand side of Equation (\ref{p-x-p}) represent different
substituting schemes as we will show in the following subsections
and in the appendix. For example, we can propose a different
arrangement $F(\xi )=e^{f_1(\xi )\hat{A}}e^{f_2(\xi )\hat{B}
}e^{f_3(\xi )\hat{A}}$, or inclusive add the commutator $\hat{C}$:
$F(\xi )=e^{h(\xi )\hat{C}}e^{f(\xi )\hat{A}%
}e^{g(\xi )\hat{B}}$.

\subsection{Application: The one-dimensional harmonic oscillator.}

One of the most important systems in Quantum Mechanics is the
harmonic oscillator. For it serves both to model many physical
systems occurring in nature and to show the analytical solution of
the Schr\"{o}dinger equation. The Schr\"{o}dinger equation for
this system has been solved in two ways, firstly by analytically
solving the eigenvalue equation and, secondly, by defining the
creation and annihilation operators \cite{book1,book2}. We solve
here the problem using the evolution operator method. This method
allows us to find the evolution for the harmonic oscillator and
avoids to deal with the stationary states.

 For the one-dimensional harmonic oscillator
the wave function at time $t$ is given by:
\begin{equation}
\Psi (x,t)=\exp \left[ -\frac{it}{\hbar }\left( \frac{\hat{p}^{2}}{2m}+\frac{%
1}{2}mw^{2}\hat{x}^{2}\right) \right] \Psi (x,0).
\label{oscillator-ini}
\end{equation}
Defining $\hat{A}=-\left( itmw^{2}/2\hbar \right) \hat{x}^{2}$, $\hat{B}%
=-\left( it/2m\hbar \right) \hat{p}^{2}$, and using the
commutation rules between $\hat{x}$ y $\hat{p}$ we can deduce the
following commutation rules:
\begin{equation}
\left[ \hat{A},\hat{B}\right] =\hat{C},\quad\left[
\hat{A},\hat{C}\right] =-2w^{2}t^{2}\hat{A}\quad and\quad\left[
\hat{B},\hat{C}\right] =2w^{2}t^{2}\hat{B},
\label{oscillator-commu}
\end{equation}
where $\hat{C}=\frac{iw^{2}t^{2}}{2\hbar }\left( \hat{x}\hat{p}+\hat{p}\hat{x%
}\right) $. If we identify $\gamma =w^{2}t^{2}$, then these
commutation relations correspond to that of Equation
(\ref{commu-oscillator}). Therefore, using the factorization found
in Eq. (\ref{p-x-p}), the Equation (\ref{oscillator-ini}) becomes:
\begin{equation}
\Psi(x,t)=e^{\mu(t)\frac{d^2}{dx^2}}e^{-\delta(t)x^2}e^{\mu(t)\frac{d^2}{dx^2}}\Psi(x,0),
\label{funct-oscillator2}
\end{equation}
where
\begin{eqnarray}
\mu(t)&=&\frac{i\hbar}{2m\omega}\tan(\omega t/2), \nonumber  \\
\delta(t)&=&\frac{im\omega}{2\hbar}\sin(\omega t). \label{mu}
\end{eqnarray}

For the one-dimensional harmonic oscillator the wave function at
time $t$ is obtained by operating with the evolution operator, i.
e. Eq. (\ref{funct-oscillator2}), on the initial wave function.
Taking as an initial wave function:
\[
\Psi (x,0)=\frac{1}{(\sigma \sqrt{\pi })^{1/2}}e^{-\frac{x^{2}}{4\sigma ^{2}%
}},
\]
 we finally obtain the state of the system at any time $t$ is:
\begin{eqnarray}
\Psi (x,t)= \frac{1}{(\sigma \sqrt{\pi})^{1/2}}
\frac{1}{\sqrt{\cos(\omega
t)+(2i\hbar/m\omega)(1/4\sigma^2)\sin(\omega t)}}
\nonumber  \\
\times \exp \left[ -\frac{(im\omega/2\hbar)\sin(\omega
t)-(1/2\sigma^2)\sin^2(\omega t/2)+(1/4\sigma^2)} {\cos(\omega
t)+(2i\hbar/m\omega)(1/4\sigma^2)\sin(\omega t)} x^{2}%
\right]. \label{psi(x,t)-oscillator}
\end{eqnarray}

From Equation (\ref{psi(x,t)-oscillator}) we can calculate the
probability distribution function:
\begin{eqnarray}
|\Psi(x,t)|^2=\frac{1}{(\sigma\sqrt{\pi})\left[\cos^2(\omega
t)+(2\hbar/m\omega)^2(1/4\sigma^2)^2\sin^2(\omega t) \right]^{1/2}} \nonumber \\
\times\exp\left[-\left(\frac{2(1/4\sigma^2)}{\cos^2(\omega
t)+(2\hbar/m\omega)^2(1/4\sigma^2)^2\sin^2(\omega t)}\right)x^2
\right]. \label{x-p-x-fin-cuadrado}
\end{eqnarray}

In the preceding case we have used the following trigonometric
identities: $1-2\sin^2(\omega t/2)=\cos(\omega t)$ and
$\sin(\omega t)=2\sin(\omega t/2)\cos(\omega t/2)$.

\subsection{Another way to factorize the harmonic oscillator}
In this subsection we present another way to factorize the
evolution operator the harmonic oscillator. Then, we apply the new
factorization on an initial wave function.

In this case we propose the factorization function as:
\begin{equation}
F(\xi )=e^{\xi (\hat{A}+\hat{B})}=e^{f_1(\xi )\hat{A}}e^{f_2(\xi )\hat{B}%
}e^{f_3(\xi )\hat{A}}. \label{x-p-x}
\end{equation}

Applying the method of factorization, we obtain the following set
of differential equation:
\begin{eqnarray}
\frac{df_2(\xi)}{d\xi }+\gamma f_2^{2}(\xi)\frac{df_3(\xi)}{d\xi
} &=&1,  \nonumber \\
\frac{df_1(\xi)}{d\xi}+\gamma f_1^2(\xi)
\frac{df_2(\xi)}{d\xi}+\frac{df_3(\xi)}{d\xi}-2\gamma f_1(\xi)
f_2(\xi) \frac{df_3(\xi)}{d\xi}+\gamma^2 f_1^2(\xi) f_2^2(\xi)
\frac{df_3(\xi)}{d\xi}&=&1,  \nonumber \\
f_1(\xi) \frac{df_2(\xi)}{d\xi
}-f_2(\xi)\frac{df_3(\xi)}{d\xi}-\gamma
f_1(\xi)f_2^2(\xi)\frac{df_3(\xi)}{d\xi}&=&0,  \label{difer-x-p-x}
\end{eqnarray}
subjected to the initial condition $F(0)=1$, which means:
$f_1(0)=f_2(0)=f_3(0)=0$. This set of differential equations is
identical to that of Eq. (\ref{difer-p-x-p}). By solving Equation
(\ref{difer-x-p-x}), we obtain the following solutions:

\begin{eqnarray}
f_1\left( \xi \right)&=&f_3\left( \xi \right)
=\frac{1}{\sqrt{\gamma}}\tan \left( \xi \sqrt{
\gamma }/2\right) ,  \\
f_2\left( \xi \right)&=&\frac{1}{\sqrt{\gamma }}\sin \left( \xi
\sqrt{\gamma } \right).
\end{eqnarray}

\subsubsection{Application}

Now, using the factorization given by Eq. (\ref{x-p-x}) to solve
the harmonic oscillator problem, we obtain the following evolution
function:
\begin{equation}
\Psi(x,t)=e^{-\alpha(t)x^2}e^{\beta(t)\frac{\partial^2}{\partial
x^2}}e^{-\alpha(t)x^2}\Psi(x,0),\label{x-p-x-fac}
\end{equation}
where
\begin{eqnarray}
\alpha(t)&=&\frac{im\omega}{2\hbar}\tan(\omega t/2), \nonumber \\
\beta(t)&=&\frac{i\hbar}{2m\omega}\sin(\omega t).
\end{eqnarray}
If we use again the initial wave function:\[
\Psi (x,0)=\frac{1}{(\sigma \sqrt{\pi })^{1/2}}e^{-\frac{x^{2}}{4\sigma ^{2}%
}},
\]
we obtain the following evolving wave function:

\begin{eqnarray}
\Psi(x,t)=\frac{1}{(\sigma\sqrt{\pi})^{1/2}\left[\cos(\omega
t)+(2i\hbar/m\omega)(1/4\sigma^2)\sin(\omega t) \right]^{1/2}} \nonumber \\
\times\exp\left[-\left(\frac{(im\omega/2\hbar)\sin(\omega
t)-(1/2\sigma^2)\sin^2(\omega t/2)+(1/4\sigma^2)}{\cos(\omega
t)+(2i\hbar/m\omega)(1/4\sigma^2)\sin(\omega t)}\right)x^2
\right]. \label{x-p-x-final}
\end{eqnarray}

Equation (\ref{x-p-x-final}) is exactly the same wave function
found in Subsection A, i. e. Equation (\ref{psi(x,t)-oscillator}).
Therefore we can conclude with one of the main points of this
paper: \textbf{\emph{The factorization could be made of different
ways and all of them have to give the same result when they are
applied to an initial wave function}}.

The evolution operator of the Equation (\ref{x-p-x-fac}) was also
proved by Beauregard \cite{bea}. However in this work was not used
the factorization method and the solution is given by an ansatz.

\begin{figure}
  \includegraphics[width=5.5in,draft=false,clip]{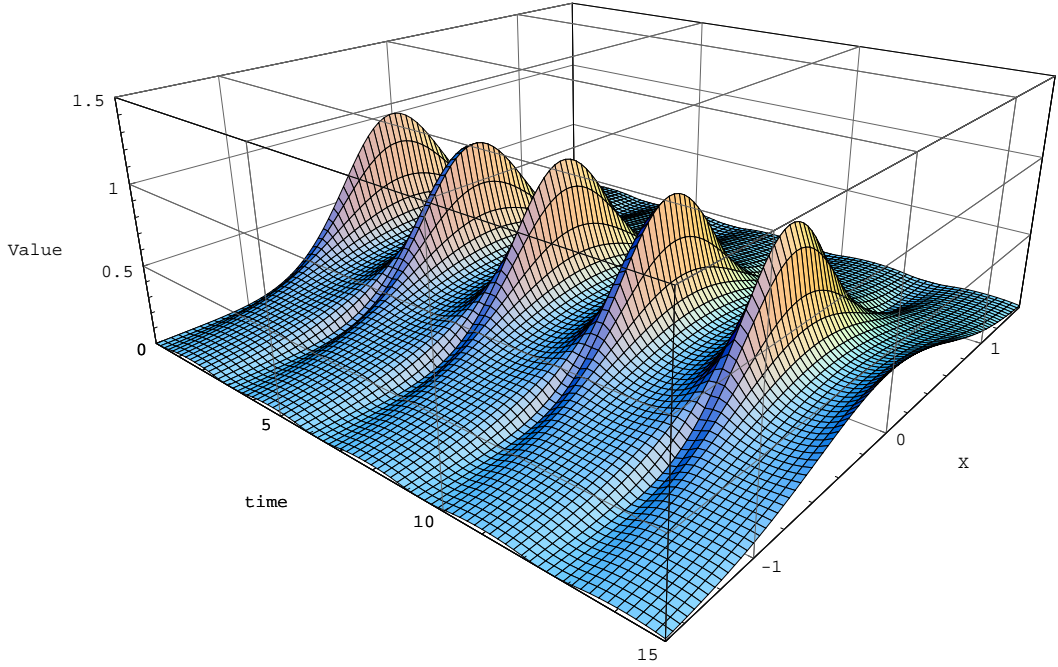}\\
  \caption{A plot of the function $|\Psi(x,t)|^2$, given by Equation (\ref{x-p-x-fin-cuadrado}).}
  \label{figx-s}
\end{figure}
In Fig. (\ref{figx-s}) we have plotted the probability density
given by Eq. (\ref{x-p-x-fin-cuadrado}). Were we have set
$\sigma_0=\sqrt{\hbar/m\omega}=1$ and $x_0=1$.

\section{A note}
Holstein and Swift published a paper in which they presented a
cautionary note about the usefulness of the evolution operator
method for obtaining the wave function at any future time $t$ from
the one at $t=0$ \cite{holstein}. Notably, in this paper Holstein
and Swift showed a particular case where the \textit{evolution
operator method} does not works, but if this case is analyzed by
the \textit{eigenstates method} it works very well. That is to
say, the results obtained with both methods do not coincide.
Therefore, a contradiction between the \textit{evolution operator
method} and the \textit{eigenstates method}  arise. the goal in
this section is to present a solution of this problem.

Firstly, we recall the arguments of reference \cite{holstein}.
 In their  argumentation, they
considered a ``free particle" represented by a one-dimensional
wave packet described by the function
\begin{eqnarray}
\psi _{H}(x,0) &=&\exp \left[ -a^{2}/\left( a^{2}-x^{2}\right) \right] ,%
\quad for \quad \left\vert x\right\vert <a \nonumber \\
&=&0,%
 \quad for \quad \left\vert x\right\vert \geq a. \label{psi_h}
\end{eqnarray}%
They argued that $\psi _{H}(x,0)$ is a ``good" function because
$\psi _{H}(x,0)$ and all its derivatives exist, are continuous for
all $x$, and vanish faster than any power as $\left\vert
x\right\vert \longrightarrow \infty $. When they apply the
evolution operator $\exp \left( -\frac{it}{\hbar }\hat{H}\right) $
to the function $\psi _{H}(x,0)$, they found that
\begin{equation}
\sum_{n=0}^{\infty }\left( \frac{i\hbar t}{2m}\right) ^{n}\left(
n!\right) ^{-1}\left( \frac{d^{2}}{dx^{2}}\right) ^{n}\psi
_{H}(x,0)=0, \label{zero}
\end{equation}
 when $\left\vert x\right\vert >a$ since
$\psi _{H}(x,0)$ and all its derivatives vanish for $\left\vert
x\right\vert
>a$. From this result their conclusion was that the particle
described by the function $\psi_H(x,0)$ is confined within
$-a<x<a$ for all time. That is to say, the wave packet does not
spread. However, if the problem is solved using the eigenstates
method then the wave packet do spread, see reference
\cite{holstein}.

In the next two subsections we analyze this argument from two
points of view. In subsection A we analyze it from the
mathematical point of view. In the subsection B we give a physical
argument.

\subsection{Mathematical view}

Mathematically, the argument given by Holstein and Swift is well
established. From the mathematical point of view, they correctly
stated that the function is a well behaved function. That is,
because outside of the interval $|x| \geq a$ it vanishes, this
function does not have any singularity and, then, it is an
analytical function. Therefore, the conclusion is that it is not
possible to apply the series of $e^{-it\hat{H}/\hbar}$ to the
function $\psi_H(x)$. In fact, there is a large set of such
function, i. e. $C^\infty$, see reference \cite{klein}.

From this conclusion we can deduce that the evolution operator
method fail when one apply it on $\psi_H(x)$. This is a very
subtle problem. As it was stated in the previous paragraph, the
function $\psi_H(x)$ is a well behaved function from the
mathematical point of view. Therefore, it seems as if the
evolution operator method fail when it is applied to an analytical
function.

However, after a careful analysis, the only that can be concluded
is that the evolution operator method fail for non analytical
function.  For example, the analysis of J. R. Klein concludes that
the evolution operator method hold for $\psi(x,0)$ in a suitable
dense subset, see section IV of reference \cite{klein}. Another
possible argument, stated in the following subsection, involves
the differences between the Hermitian and self adjoint operators
\cite{klein} and the fact that the particle is free.

Before to give a possible argument, let us recall a similar
function studied by Araujo, et. al. \cite{araujo}. Araujo, et.
al., exemplify with the function:
\begin{eqnarray}
u_{ab}=e^{1/(x-a)(x-b)}, \quad for \quad a<x<b, \nonumber \\
u_{ab}=0, \quad for \quad 0\leq x \leq a \quad and \quad x\geq b.
\label{arau}
\end{eqnarray}
For the entire interval, $-\infty<x<\infty$, this function is very
similar to the function of Holstein and Swift, $\psi_H(x,0)$.
However, Araujo, et. al., use the function (\ref{arau}) only to
show that the Hamiltonian is not a self-adjoin operator
\cite{araujo}.

\subsection{Physical view}

In this subsection, we will show that the function used by
Holstein and Swift is not a valid physical function in the case of
the free particle. A crucial point is that unbounded operators can
not be defined on all functions of the Hilbert space
\cite{klein,araujo}. In order to be able to argue this point, we
write the following two explicit assumptions given by the cited
authors \cite{holstein}:
\begin{enumerate}
\item[1)] The particle is a free particle. \item[2)] The state of
the particle at the initial time is given by the function
$\psi_H(x,0)$.
\end{enumerate}
Our principal point will be that the statements 1) and 2) can not
be true at the same time.

\begin{figure}
  \includegraphics[width=5.5in,draft=false,clip]{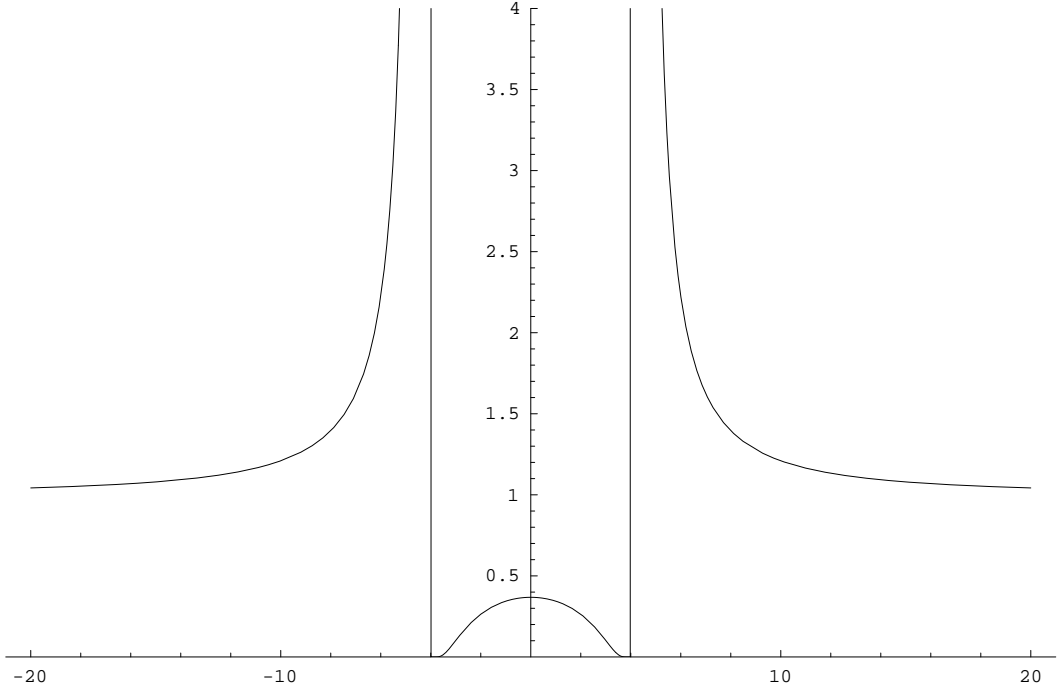}\\
  \caption{A plot of the function $\psi_{H}(x,0)$.}
  \label{fig3}
\end{figure}

We begin analyzing the function from the mathematical point of
view in the entire domain of $x$, that is $\left\{ x\in R|-\infty
\leq x\leq \infty \right\} $, see Fig. (\ref{fig3}), where the
particle can be found. Notice that the function of Holstein and
Swift is defined as zero outside the interval $|x|>a$, therefore,
strictly speaking the Fig. (\ref{fig3}) is not $\psi_H(x)$.
However, we use the entire interval $-\infty$ to $+\infty$ because
the analysis of reference \cite{holstein} is for a free particle.

This is a very peculiar function because it has two singularities
and many limits:
\begin{eqnarray}
Lim _{x\rightarrow_{-}^{+}\infty}\psi_H(x,0)=1, \nonumber \\
Lim _{x\rightarrow a^+}\psi_H(x,0)=\infty, \nonumber \\
Lim _{x\rightarrow a^-}\psi_H(x,0)=0, \nonumber \\
Lim _{x\rightarrow -a^+}\psi_H(x,0)=0, \nonumber \\
Lim _{x\rightarrow -a^-}\psi_H(x,0)=\infty. \label{limites}
\end{eqnarray}
From the previous equation we can say that the function is not
continuous for all $x$ (remember that in this subsection we are
studying it in the whole real axe). Furthermore, it does not
vanish faster than any power of $x$ as $\left\vert x\right\vert
\rightarrow \infty $. Therefore, the function is not continuous and $%
\int_{-\infty }^{\infty }\left\vert \psi (x)\right\vert ^{2}dx$ is
not finite, that is, it is not a square integrable function in the interval $%
-\infty \leq x\leq \infty $. It is important to stress that we are
analyzing the case in which the function has a definite value
outside the interval $|x|> a$.

Now, Holstein and Swift \cite{holstein} define the function as
zero in the interval $|x|> a$. With this restriction the function
is mathematically a ``good" function and can pass as a valid
function. That is, as it was stated in the previous subsection, in
this case the function does not have the behavior given in Eq.
(\ref{limites}). However, one can wonder the next questions: How
can the free particle know that it is restricted to the interval
$|x|> a$ where the behavior of the function is ``good"?, How an
state of a physical system can be represented by a function which
is mathematically defined as zero outside some interval?

At this point, let us recall the meaning of the wave function in
Quantum Mechanics. In the first place, the wave function
represents the physical state of a quantum system. That is to say,
it represents a combination of the physical properties like
energy, momentum, position, etcetera, that can be ascribed to the
system. In the second place, $\left\vert \psi (x,t)\right\vert
^{2}dx$ gives the probability that the particle could be found
between $x$ and $dx$. Then, the wave function carries the whole
information available for the system. For example, a confined
particle is restricted to have certain eigenfunctions that belong
to the domain of the Hamiltonian, and certain eigenfunctions that
belongs to the domain of the momentum operator
\cite{araujo,bonneau,capri,gieres}, see also reference
\cite{jordan}. The most severe restriction is $\int |\psi(x)|^2
dx<0$ in the entire interval.

Now, we can give a preliminary answer based in a physical insight
to the question quoted in the previous paragraph: How can the free
particle know that it is restricted to the interval $|x|\geq a$?
By definition, a free particle is a not restricted particle.
Therefore, the answer to the question is that there is no way that
the particle knows that it is restricted to certain interval, at
least if there is not an infinite well where the particle is
confined. That is, only in the case of a particle confined in an
infinite well we can set the condition $\psi(x,0)=0$ outside of
the well, and the Physics changes from that associated to a free
particle to that associated to a confined particle \cite{garba}.

As a conclusion of the previous paragraph, we can state that
physically the wave function $\psi_H(x,0)$ (when it is defined in
the entire interval $-\infty$ to $+ \infty$ ) is not valid for the
free particle. In fact, the answer is related with the differences
between Hermitian and self-adjoint operators. Mathematically, an
operator consists of a prescription of operation together with a
Hilbert space subset on where the operator is defined
\cite{klein,araujo,bonneau,gieres}. That is, the functions have to
belong to the domain of the operator, and if the operator is
defined in some interval then the set of functions where the
operator is defined, i. e. their domain, have to be defined in the
same interval. Therefore, we think that the above example is
mistake because the function $\psi_H(x,0)$ does not belong to the
domain of the hamiltonian operator of the free particle.

Let us explain, the difficulties comes from the fact that in
Quantum Mechanics the observable is represented by operators (in a
Hilbert space) and the physical states are represented by vectors
(wave functions). However, the definitions of both operators and
vectors given in most textbooks of Quantum Mechanics are very
weak. The majority of them define an operator as an action that
changes a vector in another vector, and after that they define
Hermitian operators as a symmetric operator. However, there is not
any mention about the domain of the operator and the differences
between a self-adjoint and Hermitian operators. Because of this
weak definition there are many problems or ``paradoxes" in the
calculations of physical properties, see the examples given in
references \cite{araujo,bonneau,gieres}. To handle these problems
the concept of self-adjoint extension is reviewed in references
\cite{araujo,bonneau,gieres}. Also see references
\cite{griffiths,lucio}.

 Moreover, we adhere
the recommendation of Klein \cite{klein}, Araujo et. al.
\cite{araujo}, Bonneau et. al. \cite{bonneau} and Gieres
\cite{gieres}: \textit{it is necessary to define always the domain
of the operators}. Therefore, in order to droop up all these
problems, we think that it is better to define operators in
analogy with the definition of a function. Additionally, it is
necessary to clearly state that, as the cited authors have pointed
out, unbounded operators cannot be defined on all vectors of the
Hilbert space. In particular, it has to be stated that in order
that a function can be valid as an initial state it has to be
inside of the domain of the hamiltonian $\hat{H}$. On the other
hand, the only way to set the condition that a wave function is
zero outside of some interval is to collocate an infinite well in
that interval.

Therefore, the main point in this subsection is that because the
function $\psi_H(x,0)$ is not square integrable in the interval
$-\infty< x<\infty$, then it does not belong to the domain of the
Hamiltonian operator of the free particle. Therefore the function
$\psi_H(x,0)$ can not represent a state of the free particle. This
means that the statements 1) and 2) are not true at same time.

\section{conclusion}
From the work made in this paper, we can conclude that the
evolution operator method is an efficient method to calculate the
evolution of a wave function. This method requires, at first
instance, the factorization of exponential operators. The
factorization allows us to apply the exponential operator
individually. We have shown how this method works and apply it in
elementary unidimensional cases.

As you may guess, all methods have their troubles and limitations.
One trouble of the evolution operator method is that it is not
always possible to find the factorization of the exponential
operator. Another trouble with this method is that it is not
always possible to group the evolving function in a single
expression as we show in the Appendix A. However, this method is
increasingly used by many authors. For example, we can recall the
work of Balasubramanian \cite{bala} who discussed the time
evolution operator method with time dependent Hamiltonians. Also
see reference \cite{cheng}.

\acknowledgments  We thanks the useful comments of Dr. F. A. B.
Coutinho. It is important to point out that he does not agree with
part of the content of section V. We would like to thank the
support from Consejo Nacional de Ciencia y Tecnolog\'{i}a
(CONACYT).

\appendix
\section{}
Here we show another way to factorize the harmonic oscillator. In
this case, we  define the function as:
\begin{equation}
F(\xi )=e^{\xi (\hat{A}+\hat{B})}=e^{h(\xi )\hat{C}}e^{f(\xi )\hat{A}%
}e^{g(\xi )\hat{B}}. \label{funct-oscilator}
\end{equation}
Remember that $\hat{C}=[\hat{A},\hat{B}]$. By differentiating
Equation (\ref{funct-oscilator}) with respect to $\xi $, we
obtain:
\begin{equation}
 \frac{dF(\xi )}{d\xi
}=(\hat{A}+\hat{B})F(\xi ), \label{der1-oscillator}\end{equation}
 and
\begin{equation}
\frac{dF(\xi )}{d\xi }=\left[ \left( f\frac{dg}{d\xi }+\frac{dh}{d\xi }%
\right) \hat{C}+e^{2\gamma h(\xi )}\frac{dg}{d\xi
}\hat{B}+e^{-2\gamma h(\xi )}\left( \frac{df}{d\xi }+\gamma
f^{2}\frac{dg}{d\xi }\right) \hat{A}\right] F(\xi
),\label{der-oscillator}
\end{equation}

By equating Equations (\ref{der1-oscillator}) and
(\ref{der-oscillator}), we obtain the following system of
differential equations:

\begin{eqnarray}
e^{-2\gamma h(\xi )}\left( \frac{df}{d\xi }+\gamma
f^{2}\frac{dg}{d\xi }
\right)&=&1,  \nonumber \\
e^{2\gamma h(\xi )}\frac{dg}{d\xi }&=&1,  \nonumber \\
f\frac{dg}{d\xi }+\frac{dh}{d\xi }&=&0, \label{difer-oscillator}
\end{eqnarray}
subjected to the initial condition:
\[
F(0)=1.
\]
By solving Equation (\ref{difer-oscillator}), with the initial
condition, we obtain the following solutions:

\begin{eqnarray}
f\left( \xi \right)&=&\frac{1}{\sqrt{\gamma }}\frac{\tanh \left(
\xi \sqrt{\gamma }\right) }{sech^{2}\left( -\xi \sqrt{\gamma }\right) },  \\
g\left( \xi \right)&=&\frac{1}{\sqrt{\gamma }}\tanh \left( \xi
\sqrt{\gamma }
\right) ,  \\
h\left( \xi \right)&=&\frac{1}{2\gamma }\ln \left[ sech^{2}\left(
\xi \sqrt{ \gamma }\right) \right],
\end{eqnarray}
and we obtain the factorization we were looking for.

\subsection{Aplication}

The trouble with the factorization given in Eq.
(\ref{funct-oscilator}) is that at some time we have to apply the
exponential, which contains the operator $\hat{C}$, of the form:
\begin{equation}
e^{a_1xp}e^{x^2/4\sigma^2},
\end{equation}
with $a_1$ a constant. The application of this exponential means
to apply$ \left(xp \right)^n\{e^{x^2} \}$ which produces the
following  set of polynomials $A_n(x)$ ($n=0, 1, 2, \dots$):
\begin{eqnarray}{}
 A_0&=&1 \nonumber \\
A_1&=&2x^2 \nonumber \\
A_2&=&4x^4-4x^2 \nonumber \\
A_3&=&8x^6-24x^4+8x^2 \nonumber \\
A_4&=&16x^8-96x^6-112x^4+16x^2 \\
\dots \nonumber
\end{eqnarray}

However, we were not able to find the generating function.

\end{document}